\documentstyle[aps,prb,preprint]{revtex} 
\begin{document}
\title{ \LARGE Spin state and phase competition in TbBaCo$_{2}$O$_{5.5}$ and 
the lanthanide series $Ln$BaCo$_{2}$O$_{5+\delta}$ (0$\leq\delta\leq$1) } 

\author{ Hua Wu\\}

\address{\normalsize
 Max-Planck-Institut f\"ur Physik komplexer Systeme, D-01187 Dresden, Germany\\
 and Institute of Solid State Physics, Academia Sinica,
           230031 Hefei, P. R. China\\}

\maketitle
\vspace{2cm}

  A clear physics picture of TbBaCo$_{2}$O$_{5.5}$ is revealed on the 
basis of density functional theory calculations. 
An antiferromagnetic (AFM) superexchange coupling between the 
almost high-spin Co$^{3+}$ ions competes 
with a ferromagnetic (FM) interaction mediated by both $p$-$d$ exchange and 
double exchange, being responsible 
for the observed AFM-FM transition. And the metal-insulator transition is
accompanied by an $xy$/$xz$ orbital-ordering 
transition. Moreover, this picture can 
be generalized to the whole lanthanide series, and 
it is predicted that a few room-temperature
magnetoresistance materials could be found in
$Ln$Ba$_{1-x}$$A_{x}$Co$_{2}$O$_{5+\delta}$ 
($Ln$=Ho,Er,Tm,Yb,Lu; $A$=Sr,Ca,Mg). \\


\newpage

 $Introduction.$ Materials with an interesting combination of electronic and 
magnetic properties are of current considerable interest, both from the basic 
points of view and as to their promising technological utilities. Well known 
one of this class is the colossal magnetoresistance (MR) manganite 
$R_{1-x}$$A_{x}$MnO$_{3}$ ($R$:rare earth; $A$:alkaline earth).\cite{c1}
Its FM metallicity has been long explained in terms of the double 
exchange (DE) model,\cite{c2} 
where the spin of Mn $e_{g}$ conduction electron is forced to be 
parallel to the localized $t_{2g}$ spin due to strong on-site Hund coupling, 
and the effective electron hopping of the $e_{g}$ electron is maximum when 
the neighboring Mn spins are parallel. Recently the DE 
model was modified by including a strong electron-phonon interaction
stemming from the Jahn-Teller (JT) splitting of the Mn$^{3+}$ ions.\cite{c3} 
And late it was argued$^{4,5}$ that owing to a considerable amount of 
O $2p$ holes in the 
charge-transfer$^{6}$ (CT) oxides, $p$-$d$ exchange also plays a crucial role in
determining the $T_{\rm C}$ value and transport properties.
The MR materials, especially those with a high MR value at room-$T$ and low
field have potential applications for magnetic memory and 
actuator devices. 

 Recently, the perovskite cobaltities of lanthanides 
$Ln$BaCo$_{2}$O$_{5+\delta}$ (0$\leq\delta\leq$ 1) drew much attention, because
they exhibit a few fascinating phenomena such as
a metal-insulator (M-I) transition, an AFM-FM transition, and 
a giant MR effect near the AFM-FM transition temperature 
$T_{\rm i}$.\cite{c7,c8,c9,c10} These intricate 
behaviors are closely related to an interrelation among the spin, charge, and
orbital degrees of freedom of the Co ions.
In most of the published experimental studies,\cite{c7,c8,c9}  
the abundant electronic and magnetic properties were explained in terms of
an intermediate-spin (IS) and/or a low-spin (LS) model for the Co ions. For 
example, it was suggested$^{9}$ that upon decreasing $T$ to 
$T_{\rm MI}$$\approx$340 K, TbBaCo$_{2}$O$_{5.5}$ transits from 
a Co$^{3+}$ high-spin (HS) metallic phase to an IS and lattice-distorted  
$b$-axis $3x^{2}$--$r^{2}$/$3y^{2}$--$r^{2}$ orbital-ordered (OO) 
insulating state. In this OO IS model, the observed FM state lying roughly 
between 260 and 340 K was ascribed to
both the superexchange (SE) FM along the $b$-axis and the DE FM along the 
$a$-axis, and the lower-$T$ AFM phase was ascribed to the overwhelming SE AFM. 

 However, it should be argued that such an OO IS state  
most probably leads to an AFM SE coupling along the $c$-axis as in 
the widely studied LaMnO$_{3}$.\cite{c11} 
Consequently, the above explanation of the FM state is questionable, and thus
the AFM-FM transition is somewhat hard to 
explain. In addition, there exists an 
intriguing issue$^{12}$ on the Co spin state in the isostructural 
Co$^{2+}$/Co$^{3+}$ charge-ordered 
(CO) systems YBaCo$_{2}$O$_{5}$$^{13}$ and HoBaCo$_{2}$O$_{5}$$^{14}$.  
Therefore, the Co spin state and origins of the phase transitions, being 
fundamental aspects of physics of the fascinating materials
$Ln$BaCo$_{2}$O$_{5+\delta}$, call for a further
clarification. 

  In this paper, density functional theory$^{15}$ (DFT) calculations for 
TbBaCo$_{2}$O$_{5.5}$ are exemplified to study the Co spin state and the phase
competitions. The obtained results are
distinct from the previous ones stated above, and a few fundamental 
evidences presented below clearly indicate that the 
original physics model$^{9}$ of TbBaCo$_{2}$O$_{5.5}$ is invalid. 
Instead a new physics picture is revealed,
and more importantly it can be 
generalized to the whole lanthanide series and leads to a
theoretical prediction.

 $Computational$ $details$. Used in the following calculations are the 450-K 
paramagnetic (PM) structure and 270-K distorted one of TbBaCo$_{2}$O$_{5.5}$ 
measured by neutron diffraction experiments.\cite{c9} 
Owing to a large difference between the Tb and Ba ionic radii, the O$_{4}$ 
basal squares of the alternate Co(1)O$_{6}$ octahedra and Co(2)O$_{5}$ pyramids
along the $b$-axis move towards the 
smaller-Tb layer, leading to a bent Co-O-Co bond angle along the $a$- and 
$b$-axis.

  The full-potential linearly combined atomic-orbital band method,\cite{c16} 
based on the local density approximation$^{15}$ (LDA) to DFT and on-site 
Coulomb correlation correction (LDA+$U$)\cite{c17} to the localized Tb $4f$ 
and Co $3d$ states, is adopted in the following scalar relativistic 
calculations, where no spin-orbital coupling (SOC) is included.\cite{c18} 
Tb $4f5p5d6s$, Ba $5p5d6s$, Co $3d4s$, and O $2s2p$ orbitals are treated as 
valence states. Hartree potential is 
expanded in terms of lattice harmonics up to $L$=6, and an 
exchange-correlation potential of von Barth-Hedin type is adopted.\cite{c19}
125 special $k$ points in irreducible Brillouin zone are used. 

 $Results$.  In the LDA PM band structure, the partially occupied 
Tb$^{3+}$ $4f$ bands are 
very narrow, contributing greatly to the density of states 
(DOS) at the Fermi level ($E_{F}$), as seen in Fig. 1.
This high DOS peak is to be removed by a large Mott-Hubbard splitting. 
Note that the Co $3d$ 
and O $2p$ states consist of the most concerned valence bands ranging from -8 to
3 eV, on which other electronic states have a negligible influence. 
The Co(1)/Co(2) near-degenerate $t_{2g}$ ($xz$,$yz$,$xy$) levels consist of
a subgroup of narrow bands located between --1.5 and 0.5 eV,
and the $e_{g}$ ($x^{2}$--$y^{2}$,$3z^{2}$--$r^{2}$) levels
lie, respectively, above the corresponding $t_{2g}$ ones by $\approx$1 and 
0.7 eV (the latter as in YBaCo$_{2}$O$_{5}$$^{12}$) due to a crystal-field 
(CF) splitting. The broad $e_{g}$ bands ranging from --1 to 3 eV
should be responsible for the metallic conductivity of 
TbBaCo$_{2}$O$_{5.5}$ at the high-$T$ PM phase. 

 In the LDA+$U$ calculation for the FM distorted structure, two choices of $U$=
8 and 10 eV for the Tb $4f$ electrons result in only a rigid shift of the $4f$ 
levels and have no influence on the Co $3d$ and O $2p$ states. The 
corresponding Tb$^{3+}$ ($4f^{8}$) spin moments of 5.91 and 5.94 $\mu_{B}$ 
suggest a full spin polarization.
$U$=5 eV is used for the Co $3d$ electrons,\cite{c12} and a variance of 
$\Delta$$U$=$\pm$1 eV
leads to an insignificant change of the calculated results.  
It can be seen in Fig. 2(a.1-3) that 
the Co(1)/Co(2) majority-spin $t_{2g}$ and $e_{g}$ orbitals become 
completely occupied and nearly full-filled, respectively, 
thus giving an almost HS state with an avearged Co spin moment of 3.4 $\mu_{B}$
including a total contribution of $\approx$0.4 $\mu_{B}$ from the coordinated 
oxygens. And the $e_{g}$ orbitals 
couple to the O $2p$ ones and form wide conduction bands crossing $E_{F}$.
Moreover, the near-degeneracy of the $t_{2g}$ orbitals is lifted, and
the minority-spin Co(1) $xy$ and Co(2) $xz$ orbitals are fully occupied,
giving rise to an OO of the formal Co$^{3+}$ (3$d^{6}$) ions.\cite{c20}
Furthermore, the upper valence bands (lower conduction 
bands) are mainly of the O $2p$ (Co $3d$) character, leading to a 
classification of TbBaCo$_{2}$O$_{5.5}$ as a $p$-$d$ CT oxide.\cite{c6}

  It should be noted that the above LDA+$U$ metallic solution is closely 
related to both the assumed ideal FM ordering and the itinerant O $2p$ orbital
character in this CT cobalt oxide which is something like the self-doped 
metallic ferromagnet CrO$_{2}$.\cite{c21} This result qualitatively accounts 
for a little abrupt decrease of the resistivity near $T_{\rm i}$.\cite{c9} 
While the observed insulating-like
bahavior of the so-called FM phase$^{9}$ is probably ascribed to a spin canted
structure$^{26}$ as suggested below. Moreover,the insualting ground state 
is exactly reproduced by the present AFM LDA+$U$ calculation for the 
distorted structure as seen in Fig. 2b, where the strongly suppressed 
bandwidths due to the carrier localization are evident. While the almost HS 
state and the $xy$/$xz$ OO remain unchanged.  

$Discussions$. It was theoretically found$^{12}$ that the pyramidal Co ions see
a weak CF and remain in a HS state in YBaCo$_{2}$O$_{5}$, and
it is therefore not surprising that the pyramidal Co(2) ion almost
takes the HS state in TbBaCo$_{2}$O$_{5.5}$. Also the octahedral Co(1) ion in 
TbBaCo$_{2}$O$_{5.5}$ almost takes a HS state due to a reduced CF.
It has been experimentally revealed$^{9}$ that strongly in contrast to a 
regular CoO$_{6}$ octahedron with a Co-O bond length of $\sim$1.9 \AA~, the 
cobalt and two inequivalent planar oxygens of Co(1)O$_{6}$ in the 270-K phase 
move away from the ideal CoO$_{4}$ middle plane by 0.09, 0.23 and 0.33 \AA~, 
respectively. The drastic deformation leads 
to the reduced CF splitting of $\approx$ 1 eV, and consequently
the HS state becomes energetically favorable.\cite{c22} 
Owing to a nearly complete filling of the majority-spin $e_{g}$ orbitals 
in the almost HS state, there is no freedom of the $e_{g}$ orbitals
for an OO. Instead the lattice distortion lifts the 
near-degeneracy of the $t_{2g}$ orbitals and prefers the $xy$/$xz$ OO
(see discussions in Ref. 23). These fundamental evidences 
clearly indicate that the original model$^{9}$ of the  
$3x^{2}$--$r^{2}$/$3y^{2}$--$r^{2}$ OO IS state is invalid for 
TbBaCo$_{2}$O$_{5.5}$. 
Instead it is proposed here that in this CT cobalt oxide,
the Co$^{3+}$ ions almost take a HS state with itinerant $p$-$d$ hybridized
holes of the $e_{g}$ symmetry, and they exhibit the $xy$/$xz$ OO.

  Based on this proposal, a clear physics picture of TbBaCo$_{2}$O$_{5.5}$ is 
presented.  A SE interaction (Ref. 24) between the almost HS Co$^{3+}$ ions 
gives an AFM coupling. Owing to a significant O $2p$ orbital character of the 
$p$-$d$ hybridized holes, the itinerant O $2p$ holes couple 
antiferromagnetically to the Co spins and align them ferromagnetically via the 
$p$-$d$ exchange.\cite{c4,c25} Also the itinerant Co $e_{g}$ holes 
mediate the FM coupling via DE.\cite{c2} In the low-$T$ distorted structure 
of TbBaCo$_{2}$O$_{5.5}$, the OO is stabilized,
and the $p$-$d$ holes are strongly localized due to a polaron effect and thus 
the FM coupling is disfavored, and the AFM SE coupling is overwhelming 
instead. These effects account for the observed insulating AFM ground 
state.\cite{c9} With increasing $T$ below $T_{\rm MI}\approx$340 K, the 
$p$-$d$ holes delocalize progressively but the OO survives as shown 
in Fig. 2a. As a result, a FM coupling becomes pronounced but the AFM coupling
is inherent. Thus a spin canted structure$^{26}$ (rather than the so-called
FM state$^{9}$) appears in a relatively narrow range of $T$, which is 
considered to be the reason for the high resistivity 
with a little sudden drop near $T_{\rm i}$.\cite{c9} As the $T$ increases 
further, the PM phase appears, the OO collapses, and the itinerant 
$p$-$d$ holes result in a metallic behavior. 

  The present picture should be also applicable to other intermediate 
lanthanides of this class, in view of a negligible influence of the strongly 
localized Tb$^{3+}$ $4f$ electrons on the Co and O states as shown above.
A relevant experimental evidence is that similar AFM-FM and M-I transitions 
were observed in the intermediate $Ln$BaCo$_{2}$O$_{5+\delta}$
($Ln$=Sm,Eu,Gd,Tb,Dy; $\delta\sim$0.5).\cite{c7} And in the intermediate 
lanthanides at $\delta$=1, a complete apical-oxygen occupation makes Co ions
all octahedrally coordinated and thus to behave like the octahedral Co(1) ions
in TbBaCo$_{2}$O$_{5.5}$. As a result, they 
show a great similarity to those at $\delta\sim$0.5 
in the electronic and magnetic properties.\cite{c8,c7} 

  Moreover, this picture can be generalized to the whole lanthanide series 
as discussed below, by taking 
into account variances of both the spin state and the hole behavior,
and a possible CO transition. 

  In the larger lanthanides, $e.$$g.$, $Ln$=La, Pr, and Nd,
the decreasing difference between the $Ln^{3+}$ and Ba$^{2+}$ sizes 
leads to a relaxation of the CoO$_{4}$ basal square close to the ideal middle
plane and therefore a weaker deformation of the Co-O polyhedra. As a result, 
the CF is enhanced and the Co HS state is disfavored. In particular, an IS
state is most probably stabilized in LaBaCo$_{2}$O$_{6}$ 
(as in SrCoO$_{3}$ and LaSrCo$_{2}$O$_{6}$\cite{c25}) by a strong $p$-$d$
hybridization which is responsible for its FM metallicity$^{8}$ via the 
$p$-$d$ exchange and DE mechanisms. In addition, an IS-HS mixing state could
appear in PrBaCo$_{2}$O$_{5+\delta}$ and 
NdBaCo$_{2}$O$_{5+\delta}$. Thus, the weak magnetization and semiconductive
behavior observed in them$^{8}$ can be reasonably ascribed to
the coexisting IS FM metallicity and HS AFM insulating behavior.\cite{c26}

  In contrast, a larger CoO$_{4}$ displacement away from the middle plane is 
expected in the smaller lanthanides, $e.$$g.$, HoBaCo$_{2}$O$_{5}$, as well as
YBaCo$_{2}$O$_{5}$ (See Refs. 9 and 13 for comparison). Consequently, the CF 
is strongly 
reduced, and the Co ions take a HS state and give a strong AFM SE coupling. 
The $p$-$d$ holes are localized due to the strongly bent Co-O-Co bonds. Thus, 
these smaller lanthanides should exhibit a pronounced AFM insulating behavior 
with high $T_{\rm N}$
as observed in HoBaCo$_{2}$O$_{5}$ ($T_{\rm N}\sim$340 K)$^{14}$
and YBaCo$_{2}$O$_{5}$ ($T_{\rm N}\sim$330 K).\cite{c13} 

  Note that a CO transition 
is more likely realized in the intermediate and smaller lanthanides  
($\delta$=0,1) with mixed-valence Co ions 
as observed in HoBaCo$_{2}$O$_{5}$$^{14}$ and YBaCo$_{2}$O$_{5}$$^{13}$, 
owing to the bent Co-O-Co bond angles and hence the reduced 
electron hoppings.\cite{c10,c12} This CO transition should be rather 
responsible for the M-I transition observed in the intermediate 
$Ln$BaCo$_{2}$O$_{6}$.\cite{c10,c8}

  As discussed above, with increasing $T$ the intermediate 
$Ln$BaCo$_{2}$O$_{5+\delta}$ ($\delta\sim$0.5,=1)
transit from an AFM insulator to 
a PM metal via an intermediate so-called FM state with a little sudden drop of 
high resistivity near $T_{\rm i}$. An 
applied magnetic field will inhibit the AFM ordering of the Co spins but 
favor the FM coupling (and thus enhance the $p$-$d$ hole mobility),
as was indicated by experiments$^{8}$ that 
an external magnetic field lowers the $T_{\rm i}$ and makes broader the
range of $T$ for the FM state. As a result, the magnetic field
changes the hole mobility in the intermediate lanthanides near 
the AFM-FM phase boundary more considerably than near the FM-PM or M-I phase
boundaries, giving a giant negative MR near the $T_{\rm i}$ rather than near 
the $T_{\rm C}$ or $T_{\rm MI}$.\cite{c7,c8}
In contrast, the MR effect becomes weaker in the larger lanthanides which behave
like a FM (semi)metal, and it disappears even in the smaller ones being almost 
an AFM insulator.\cite{c7}      
  
  Since a giant MR effect of the intermediate lanthanides appears near 
$T_{\rm i}$,
$e.$$g.$, 260K for TbBaCo$_{2}$O$_{6}$,\cite{c8} some lanthanides of this class
with a higher $T_{\rm i}$ value could be candidate
room-$T$ MR materials. In the sense, the smaller lanthanides, although 
intrinsically insulating, seem preferable through a modulation of the $p$-$d$
hole behavior. It can be reasonably postulated that a smaller-$A$ 
($A$=Sr,Ca,Mg) substitution for the Ba composition in
the smaller lanthanides is a promising choice. Thus, the decreasing difference 
between the $Ln$ and $A$ sizes weakens the deformation
of the Co-O polyhedra, and it leads to 
an enhancement of the $p$-$d$ hole itineracy at the expense 
of a litter drop of the high $T_{\rm i}$ value. As a result, a few 
room-$T$ MR materials could be found in 
$Ln$Ba$_{1-x}$$A_{x}$Co$_{2}$O$_{5+\delta}$
($Ln$=Ho,Er,Tm,Yb,Lu; $A$=Sr,Ca,Mg) with a possibly 
optimized combination of both the $T_{\rm i}$ value and the 
$p$-$d$ hole itineracy. Moreover, such a doping in the intermediate lanthanides 
could increase their MR values.  

  $Conclusion$. Based on LDA and LDA+$U$ calculations for TbBaCo$_{2}$O$_{5.5}$,
it has been proposed that the Co ions take an almost HS state with itinerant 
$p$-$d$ hybridized holes in the intermediate lanthanides 
$Ln$BaCo$_{2}$O$_{5+\delta}$ ($Ln$=Sm,Eu,Gd,Tb,Dy; $\delta\sim$0.5,=1), and
most likely an IS state or an IS-HS mixing one in the larger ones 
($Ln$=La,Pr,Nd; $\delta\sim$0.5,=1)
stabilized by a stronger $p$-$d$ hybridization, and a complete HS state with
localized holes in the smaller $Ln$BaCo$_{2}$O$_{5}$, $e.$$g.$ 
HoBaCo$_{2}$O$_{5}$. Thus a general physics picture of the lanthanide series   
has been revealed that a competition between the HS AFM SE 
coupling and the FM interaction mediated by both $p$-$d$ exchange and DE due to
the itinerant $p$-$d$ holes, and possible OO and/or CO transitions
related to a lattice distortion, are responsible for the FM behavior 
of the larger lanthanides, and AFM-FM and M-I transitions in the  
intermediate ones, and the AFM insulating property of the smaller ones.
Moreover, it is theoretically predicted that a few room-$T$ MR 
materials could be found in 
$Ln$Ba$_{1-x}$$A_{x}$Co$_{2}$O$_{5+\delta}$
($Ln$=Ho,Er,Tm,Yb,Lu; $A$=Sr,Ca,Mg).\ 

  $Acknowledgments$. The author thanks A.-m Hu and A. Yaresko for their 
valuable discussions. This work was cosponsored by Max Planck 
Society of Germany and National Natural Science Foundation of China.\\

\newpage
\begin{figure}

\caption{ The LDA PM DOS. Fermi level is set at zero.  
The wide $e_{g}$ bands and narrow $t_{2g}$ ones are evident. \\}

\caption{ (a.1-3) The LDA+$U$ Co $3d$ and O (Ref. 9) $2p$ DOS for the FM 
state. The solid (dashed) line denotes the majority (minority) spin. 
The almost HS and $xy$/$xz$ OO state is evident, as well as itinerant $p$-$d$
holes. (b) Partial Co $3d$ DOS (for saving space) of the LDA+$U$ insulating 
state with $b$- and $c$-axis AFM ordering. The bandwidths
will be further suppressed if the $a$-axis AFM ordering is also considered. \\}

\end{figure}

\begin{references}
\bibitem{c18}  Owing to a strong localization of the Tb $4f$ electron,  
its relativistic effect has a negligible influence on the most 
concerned Co $3d$ and O $2p$ states. In addition, a negligence of the 
Co SOC effect does not affect the present discussions on the Co spin state and 
orbital structure, since the strength of the SOC being commonly 0.01-0.1 eV  
is much weaker than the CF and Hund coupling.
\end{references}
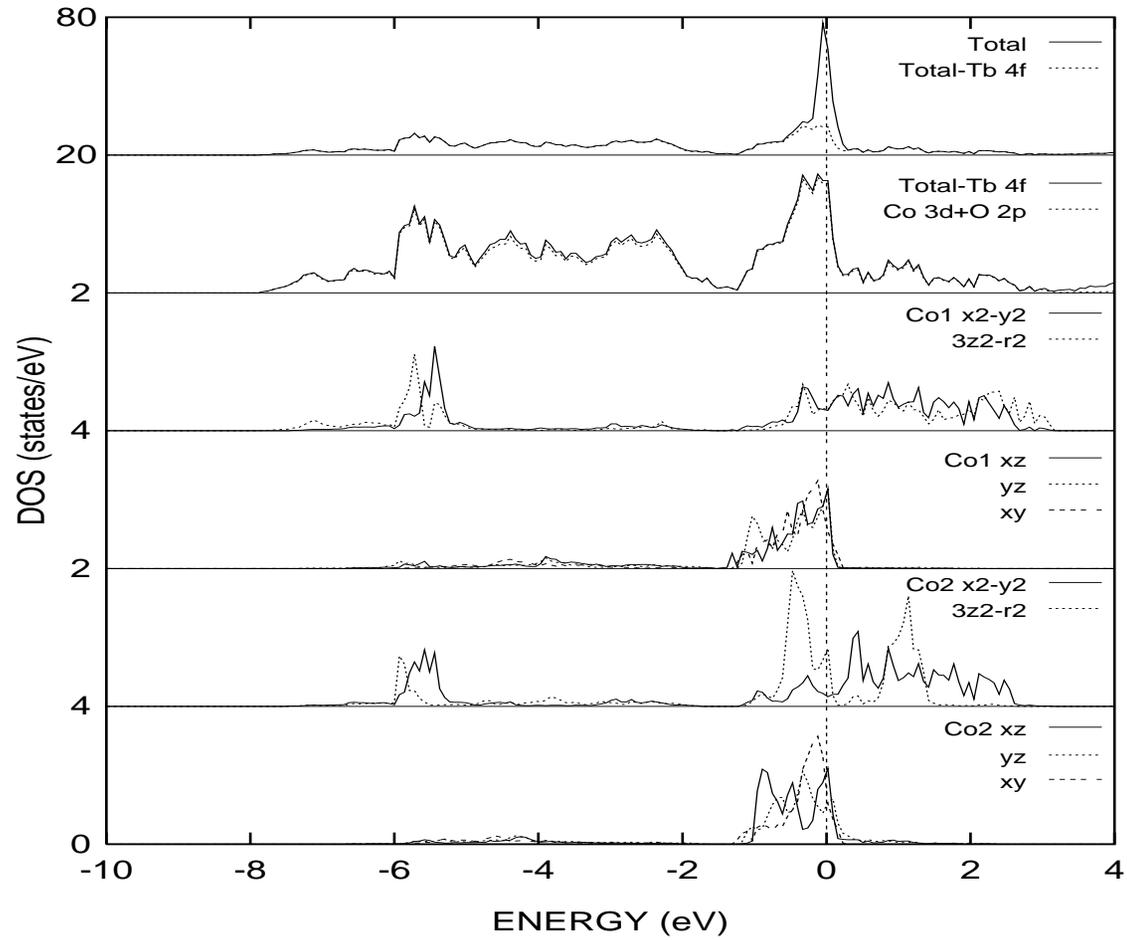
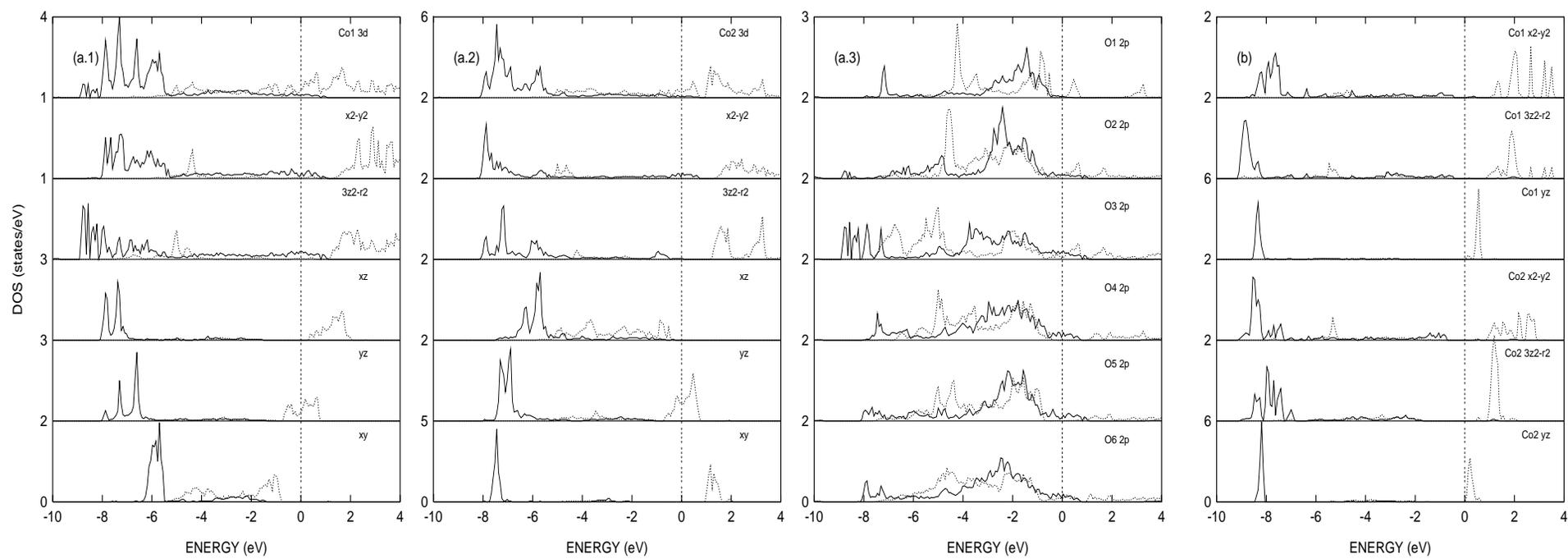
\end{document}